\documentclass[10pt,conference]{IEEEtran}
\usepackage{cite}
\usepackage{amsmath,amssymb,amsfonts}
\usepackage{algorithmic}
\usepackage{graphicx}
\usepackage{textcomp}
\usepackage[dvipsnames]{xcolor}
\usepackage[hyphens]{url}
\usepackage{fancyhdr}
\usepackage{hyperref}
\usepackage{xspace}
\usepackage{tikz}
\usepackage{multicol}
\usepackage{soul}

\usepackage{booktabs}
\usepackage{multirow}

\newcommand{\hpcayear}{2027}

\ifx\figurename\undefined \def\figurename{Figure}\fi
\renewcommand{\figurename}{Fig.}
\renewcommand{\paragraph}[1]{\textbf{#1} }
\newcommand{\subparagraph}[1]{\underline{\textit{#1}} }

\newcommand{\Sect}[1]{Sec.~\ref{#1}}
\newcommand{\Fig}[1]{Fig.~\ref{#1}}

\newcommand{\Review}[1]{}

\def\cF{{\mathcal{F}}}

\def\cS{{\mathcal{S}}}

\newcommand{\mode}[1]{\underline{\textsc{#1}}\xspace}

\newcommand{\proj}{\textsc{Zora}\xspace}

\newcommand{\PBox}[1]{\vspace*{.05cm}\noindent\fbox{\parbox{.98\columnwidth}{\vspace*{.05cm}{#1}}}\vspace*{.05cm}}

\graphicspath{{figs/}}

\title{Efficient VLA Inference via Retrieval-Based Speculation and Hardware Acceleration}

\def\hpcacameraready{} 

\author{
    \IEEEauthorblockN{Zhengyang Cheng\textsuperscript{1,\dag}, He Zhu\textsuperscript{1,\dag}, Zeyu Guo\textsuperscript{1}, Zihan Liu\textsuperscript{1, 2}, Anbang Wu\textsuperscript{1} \\
    Yiming Gan\textsuperscript{3}, Jingwen Leng\textsuperscript{1, 2}, Yu Feng\textsuperscript{1, 2, *}, Minyi Guo\textsuperscript{1, 2}}
    
    \IEEEauthorblockA{
        \textsuperscript{1}Shanghai Jiao Tong University, \textsuperscript{2}Shanghai Qi Zhi Institute \\
        \textsuperscript{3}Institute of Computing Technology, Chinese Academy of Science}
    \IEEEauthorblockA{
        $\dagger$Equal contribution, *Corresponding author 
    }
}

\fancypagestyle{camerareadyfirstpage}{%
  \fancyhead{}
  
  \fancyhead[C]{
    \ifdefined\aeopen
    \parbox[][12mm][t]{13.5cm}{\hpcayear{} IEEE International Symposium on High-Performance Computer Architecture (HPCA)}    
    \else
      \ifdefined\aereviewed
      \parbox[][12mm][t]{13.5cm}{\hpcayear{} IEEE International Symposium on High-Performance Computer Architecture (HPCA)}
      \else
      \ifdefined\aereproduced
      \parbox[][12mm][t]{13.5cm}{\hpcayear{} IEEE International Symposium on High-Performance Computer Architecture (HPCA)}
      \else
      \parbox[][0mm][t]{13.5cm}{\hpcayear{} IEEE International Symposium on High-Performance Computer Architecture (HPCA)}
    \fi 
    \fi 
    \fi 
    \ifdefined\aeopen 
      \includegraphics[width=12mm,height=12mm]{ae-badges/open-research-objects.pdf}
    \fi 
    \ifdefined\aereviewed
      \includegraphics[width=12mm,height=12mm]{ae-badges/research-objects-reviewed.pdf}
    \fi 
    \ifdefined\aereproduced
      \includegraphics[width=12mm,height=12mm]{ae-badges/results-reproduced.pdf}
    \fi
  }
  \fancyfoot[C]{}
}
\ifdefined\hpcacameraready 
\else
\fi

\ifdefined\eaopen

\fi

\begin{document}

\maketitle

\begin{abstract}

Vision-language-action (VLA) models enable robots to understand and execute complex instructions, but their autoregressive (AR) action generation introduces significant latency on resource-constrained robotic platforms.
Speculative decoding can accelerate AR-based inference by parallelizing action generation and verification; however, existing approaches rely on costly draft models, which are inefficient on edge platforms.

In this paper, we propose \proj, a hardware-software co-design framework for accelerating AR-based VLA inference.
Specifically, \proj introduces a lightweight retrieval-augmented generation (RAG)-based draft model that exploits robotic dynamics without additional neural draft models.
We further reduce batch verification overhead through a kinematics-guided action-acceptance strategy and a KV-cache-sharing attention mechanism.
To address the retrieval overhead introduced by RAG-based drafting, we co-design dedicated hardware support, \proj core, to address the memory irregularity in RAG-based drafting.
We evaluate \proj on four VLA models with the LIBERO benchmark.
\proj retains comparable accuracy while achieving up to 3.4$\times$ speedup with software-only optimization and up to 7.6$\times$ speedup with hardware support over GPUs.

\end{abstract}


\section{Introduction}
\label{sec:intro}

Embodied AI is becoming one of the foundations for next-generation intelligent systems~\cite{feng2025embodied, sun2024comprehensive}.
By integrating perception, language understanding, and action generation, embodied AI enables robots and autonomous machines to interact with the physical world.
This capability has attracted growing commercial interest, with the global humanoid robot market estimated at USD 1.55 billion in 2024 and projected to reach USD 4.04 billion by 2030~\cite{robotmarket}.
As embodied AI moves from research prototypes to real-world deployment, vision-language-action (VLA) models have become the key to allowing robots to understand complex instructions.

Mainstream VLA models~\cite{kim24openvla, pertsch2025fast, wu2024unleashing, li2024vision, kim2025fine} commonly adopt an autoregressive (AR) inference process, similar to large language model (LLM) decoding.
Instead of generating text tokens, AR-based VLA models generate action tokens sequentially and use each generated action to predict the next one.
This sequential decoding process introduces high latency on resource-constrained robotic platforms.
For example, existing AR-based algorithms barely achieve 10.0~FPS (\Fig{fig:vla_latency}) on Nvidia Jetson Thor~\cite{thorsoc}, far below the real-time requirement, 60~FPS, for robotics~\cite{guo2024prediction, li2025unified, belkhale20rth, liu2024rdt, black2026pi, kim24openvla}.
This paper thus focuses on accelerating AR-based VLA inference.

\paragraph{Issues.} 
The main bottleneck of AR-based VLA models is the decoding stage, which accounts for 69.5\% of the execution time, on average (\Fig{fig:exec_breakdown}).
The fundamental issue in decoding is that \textit{AR-based action generation is inherently sequential}.
Since each token generation depends on previous tokens, the next token decoding cannot start until the current one finishes.
To break this dependency, speculative decoding~\cite{chen2023accelerating, leviathan2023fast, miao2024specinfer} provides an opportunity to expose token-level parallelism during action decoding.
Specifically, a lightweight draft model first predicts a sequence of future actions, and the original VLA model then verifies these drafted actions in batch.
Thus, we can reduce the number of sequential decoding steps.

However, directly applying speculative decoding to robotic platforms does not necessarily guarantee great performance gains.
Conventional speculative decoding relies on sufficient compute resources on server GPUs to predict multiple drafted tokens and verify them in parallel.
In contrast, robotic platforms have limited resources.
On such platforms, verifying too many drafted tokens in a batch may not translate into proportional parallel execution.
Therefore, speculative decoding for VLA inference must explicitly account for both the cost of token drafting and the efficiency of batched verification, rather than relying solely on token-level parallelism.

\begin{figure*}
    \centering
    \includegraphics[width=\textwidth]{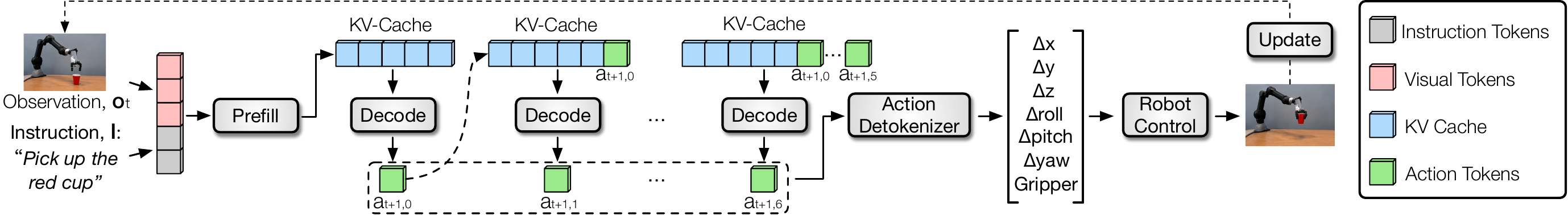}
    \caption{Inference process of single-action autoregressive VLA models. 
    With the visual observation $\mathbf{o}_t$ and language instruction $\mathbf{I}$, the model first performs a prefill step to encode the input context and generate the KV cache. During decoding, action tokens are generated autoregressively and appended to the KV cache. 
    Once all action tokens are generated, including relative movement $\Delta x_{t+1}$, rotation change $\Delta r_{t+1}$, and gripper state $g_{t+1}$, the VLA model then uses a small action detokenizer to decode the action and control the robot.
    Lastly, the execution updates the next observation for subsequent action generation.}
    \label{fig:ar_vla}
\end{figure*}

\paragraph{Framework.}
To address the challenges, we propose \proj, a unified acceleration framework for AR-based VLA on robotic platforms (\Sect{sec:algo}).
Overall, we propose three contributions to reduce the cost of drafting and verification.

First, to reduce drafting cost, we propose a lightweight drafting method that exploits the temporal regularity of robotic dynamics.
Instead of using a smaller LLM, we design a lightweight retrieval-augmented generation (RAG)-based technique to generate future action candidates.
Thus, our framework retrieves similar execution contexts from the database and uses their future actions as drafts, with low overhead.

Second, we design an action-acceptance strategy to bypass some action verifications.
Instead of verifying every drafted action via the original VLA model, \proj estimates the confidence of each drafted action based on the robotic dynamics and the consensus of multiple drafted tokens.
The high-confidence drafted action can be directly accepted without invoking batched verification.
Only low-confidence drafted actions are sent to the original VLA model for verification.

Third, we propose a KV-cache-sharing attention mechanism to further reduce the batch verification cost for the remaining actions that still require verification.
Instead of treating each drafted action as an independent sequence, our mechanism enables multiple drafted action candidates to share the same visual and language tokens during verification.
Thus, we reduce KV-cache loading overhead in attention computation.

\paragraph{Architectural Support.}
While our \proj framework substantially reduces the model verification cost, the RAG-based draft model itself becomes a new bottleneck (\Fig{fig:issue}).
Specifically, it relies on graph traversal, which introduces irregular computation.
To address this bottleneck, we design a dedicated \proj core in \Sect{sec:arch} that supports efficient graph traversal by exploiting node-level and embedding-level parallelism in RAG-based data retrieval.
Meanwhile, we introduce a dedicated checker unit to further support the kinematic and consistency checks.
These architectural augmentations provide minimal but principled hardware support for the remaining bottlenecks in our framework after algorithmic acceleration.

\paragraph{Results.} 
We implement the \proj hardware prototype and evaluate it over four VLA models~\cite{kim24openvla, wang2025unified, pertsch2025fast, goyal2025vla0}.
On the LIBERO dataset~\cite{liu2023libero}, we show that \proj achieves a comparable task success rate with no accuracy degradation. 
With our software-only optimization, \proj achieves up to 3.4$\times$ speedup on the off-the-shelf GPUs.
With our hardware support, \proj achieves up to 7.6$\times$ and 4.8$\times$ speedup compared with mobile GPUs and dedicated accelerators, respectively. 
This paper makes the following contributions:
\begin{itemize}
\item We quantitatively analyze the workload of AR-based VLA models and identify their common bottlenecks.
\item We leverage speculative decoding and robotic kinematics to reduce the workload and latency of VLA inference.
\item We propose a minimal yet principled hardware augmentation to address the irregularity of retrieval-based drafting.
\item Our co-design achieves up to 7.6$\times$ speedup over prior methods with no accuracy degradation.
\end{itemize}
\section{Background}
\label{sec:bg}

This section gives the background of AR-based VLA models (\Sect{sec:bg:vla}) and retrieval-augmented generation (\Sect{sec:bg:rag}).

\subsection{Autoregressive-Based VLA Models}
\label{sec:bg:vla}

\paragraph{VLA Model.} 
The goal of a VLA model is to translate multimodal observations into executable actions for embodied agents.
A typical VLA model takes three pieces of information as input: the current visual observation $ \mathbf{o}_t$, a language instruction $\mathbf{I}$, and the current robot state $\mathbf{s}_t$ at a given timestep $t$.
Based on these inputs, the VLA model predicts the next action $\mathbf{a}_{t+1}$ or a future action sequence, $\{\mathbf{a}_{t+1}, ..., \mathbf{a}_{t+w}\}$, where $w$ is the length of the predicted action sequence.

\paragraph{Workflow.}
Mainstream VLA models~\cite{kim24openvla, pertsch2025fast, wu2024unleashing, li2024vision, wang2025unified, goyal2025vla0} often follow an autoregressive generation paradigm, similar to LLMs.
As shown in \Fig{fig:ar_vla}, at each timestep $t$, the VLA model observes the current environment $\mathbf{o}_t$ and generates the next action $\mathbf{a}_{t+1}$.
The low-level control logic in the robot then consumes this action and moves to a new state $\mathbf{s}_{t+1}$.
The VLA model repeats this process until this task is completed.

During inference, a VLA model does not predict one action $\mathbf{a}_{t+1}$ in one decoding step.
Instead, each action is represented by multiple tokens and generated autoregressively.
E.g., a robot-arm action can be represented by seven tokens in \Fig{fig:ar_vla}.
The first three tokens encode the displacement difference $\Delta x_{t+1}$, the next three tokens encode the rotation difference $\Delta r_{t+1}$, and the last token encodes the gripper state $g_{t+1}$.
Different VLAs can have different encoding schemes~\cite{yu2025survey}.

At each control step, the VLA model first takes the instruction $\mathbf{I}$ and the current sensory observation $\mathbf{o}_t$ to perform the prefill stage, producing hidden states and the KV cache.
Then, during the decoding stage, the model generates one token at a time, e.g., $a_{t+1,0}$.
Each generated token is appended to the input token sequence and used to predict the next token, e.g., $a_{t+1,1}$.
This process repeats until all tokens of the current action are generated.
The generated action tokens are then passed to an action detokenizer, which converts the tokens into robot control commands.
Finally, these commands are fed to low-level control algorithms~\cite{williams2016aggressive, stellato2018osqp, schubiger2020gpu} to drive the robot.


\subsection{Retrieval-Augmented Generation (RAG)}
\label{sec:bg:rag}

\begin{figure}[t]
    \centering
    \includegraphics[width=\columnwidth]{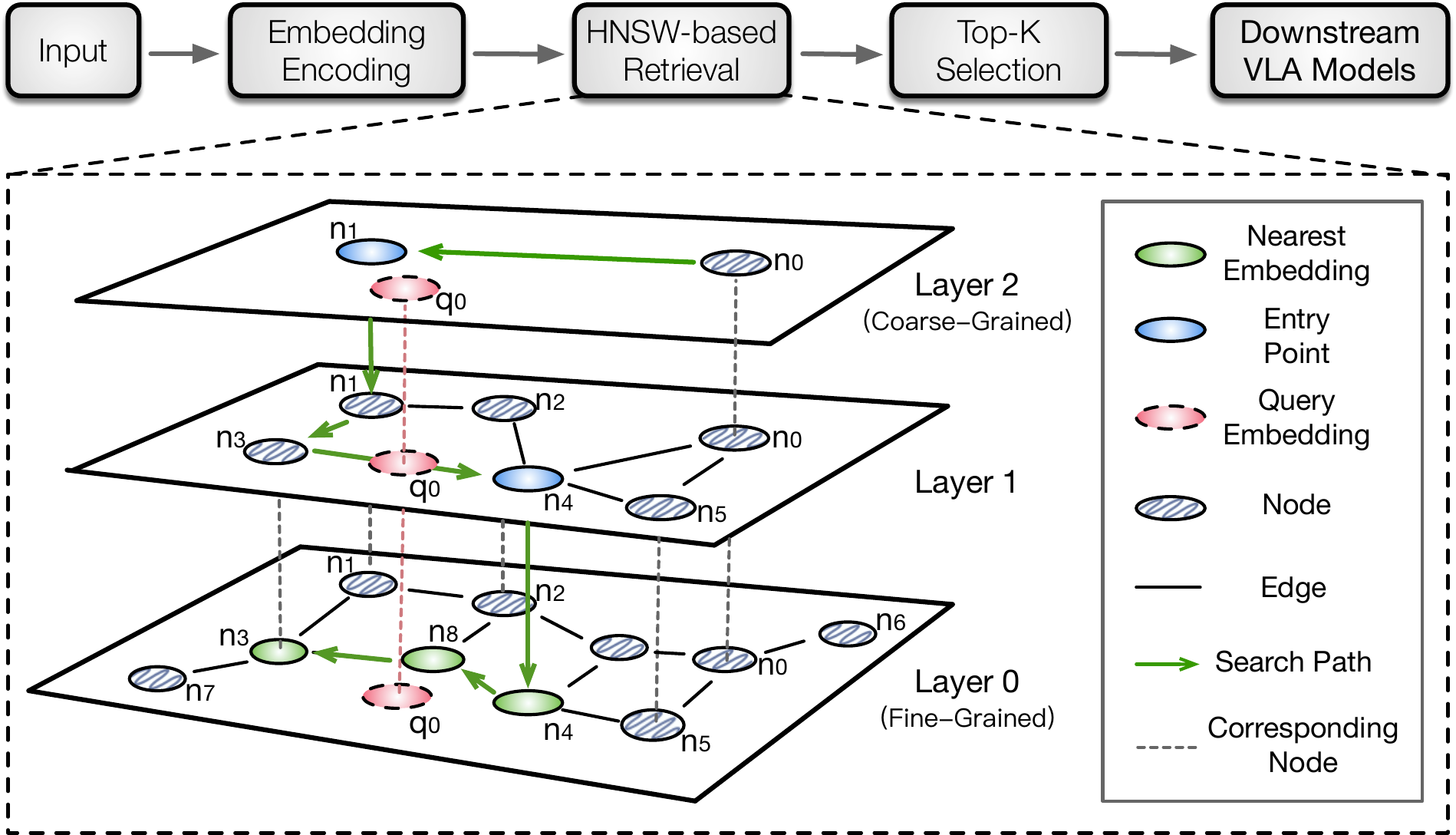}
    \caption{The HNSW-based retrieval process.
    The input is first encoded into a query embedding, which is used to search the vector database before selecting the top-$K$ retrieved results for downstream VLA models. The search database is implemented with a HNSW graph, where search proceeds in a coarse-to-fine manner, from the top layer to the bottom layer.}
    \label{fig:hnsw_process}
\end{figure}

RAG has been widely used to enhance generative models in many domains~\cite{quinn2025accelerating, jiang2025rago, shen2025hermes, liu2025heterrag, mahapatra2025in}, such as LLMs.
Its core idea is to augment model generation with relevant information retrieved from an external database, rather than relying solely on knowledge encoded in model parameters.
The basic workflow is as follows.

\paragraph{Overview.}
A RAG system consists of two stages: an offline indexing stage and an online retrieval stage.
During offline indexing, the system builds the external database with domain knowledge.
In the case of VLA, such knowledge can include robot states or trajectory data.
Each entry in this database is a key-value pair for retrieval.
The retrieval key is an embedding generated from the input query, while the retrieval value could be its associated data.
During online inference, the input query is first encoded into a query embedding.
The RAG system then searches for embeddings that are similar to the query embedding and retrieves the corresponding values.

In the VLA setting, the retrieval key can be a combined embedding of the visual observation $ \mathbf{o}_t$, a language instruction $\mathbf{I}$, and the current robot state $\mathbf{s}_t$.
The retrieval value can be the future action trajectory, $\{\mathbf{a}_{t+1}, \ldots, \mathbf{a}_{t+w}\}$, observed under a similar execution context in the RAG database.
Mathematically, each database entry can be represented as,
\begin{equation}
\mathrm{Key} = \cF_{RAG}(\mathbf{o}_t, \mathbf{I}, \mathbf{s}_t)\ \text{and}\ \mathrm{Value} = \{\mathbf{a}_{t+1}, \ldots, \mathbf{a}_{t+w}\}, \nonumber
\end{equation}
where $\cF_{RAG}$ is the embedding function and $w$ is the prediction window.
During inference, we search for contexts and their future action trajectories similar to the current robot context.

\paragraph{Hierarchical Navigable Small World.}
RAG systems often rely on approximate nearest-neighbor search to accelerate data retrieval.
One widely used algorithm in RAG systems is \textit{hierarchical navigable small world} (HNSW)~\cite{malkov2020efficient}, which organizes embeddings into a multi-layer proximity graph.

HNSW algorithm first builds a hierarchical graph with multiple layers offline.
In \Fig{fig:hnsw_process}, we use three layers as an example.
In this graph, the upper layers contain fewer nodes and provide long-range shortcuts across the embedding space, while lower layers contain more nodes and provide fine-grained nearest-neighbor search.
Each node represents an embedding and connects to a small set of nearby embeddings.
This hierarchical graph allows the search process to quickly navigate toward the region of the embedding space that is close to the query.
Since the hierarchical graph is built offline, this paper only focuses on the online search of HNSW.

\Fig{fig:hnsw_process} shows an example of the retrieval process in HNSW.
Here, the pink point $q_0$ denotes the query embedding.
The search starts from any node at the top layer (Layer 2).
At each layer, it greedily moves from the current node to a neighboring node if that neighbor is closer to the query.
This greedy search continues until no neighbor in the current layer is closer than the current node.
For instance, node $n_1$ is the nearest node to $q_0$.
This node is then the entry point for the next lower layer.

The same greedy procedure is repeated layer by layer.
In \Fig{fig:hnsw_process}, the blue nodes denote the entry points to lower layers, and the green arrows show the search path.
After reaching the bottom layer, i.e., Layer~0, HNSW performs the most fine-grained search over a denser set of nodes. 
The search continues expanding nearby candidates until it finds the closest embeddings to the query.
In this case, we need to find the top-3 nearest neighbors, shown as the green nodes, i.e., $n_3$, $n_4$, and $n_8$. 
These retrieved nearest neighbor embeddings are then passed to the downstream VLA model.
\section{Motivation}
\label{sec:motiv}

To understand the underlying performance bottlenecks of VLA models, we characterize various AR-based VLA models on a Nvidia Thor SoC~\cite{thorsoc} and show three key insights that motivate the design of our acceleration framework.

\begin{figure}[t]
\centering
\begin{minipage}[t]{0.48\columnwidth}
  \centering
  \includegraphics[width=\columnwidth]{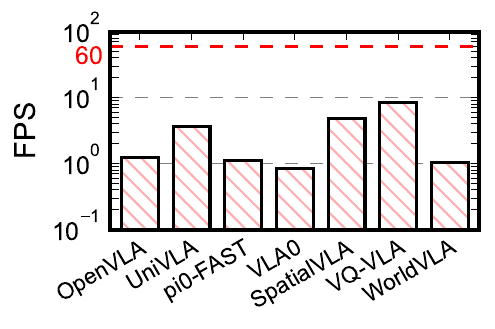}
  \caption{The frame rate of various AR-based VLA models on Thor SoC~\cite{thorsoc}. The red line highlights the real-time requirement. All models are far from real-time execution.}
  \label{fig:vla_latency}
\end{minipage}
\hspace{2pt}
\begin{minipage}[t]{0.48\columnwidth}
  \centering
  \includegraphics[width=\columnwidth]{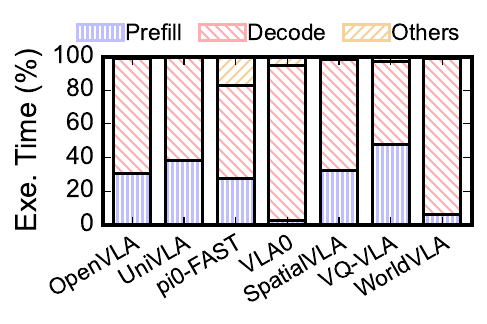}
  \caption{Execution breakdown of seven representative AR-based VLA models~\cite{kim24openvla, wang2025unified, pertsch2025fast, goyal2025vla0, qu2025spatialvla, wang25vqvla, cen2025worldvla}. The decoding stage dominates the execution time.}
  \label{fig:exec_breakdown}
\end{minipage}
\end{figure}

\vspace{1pt}
\PBox{\textbf{Insight 1:} \textit{Existing AR-based VLA models are far from real-time execution on existing mobile platforms.}~\cite{kim2025fine, zhou2026characterizing}}
\vspace{0pt}

Real-time VLA requires generating actions at over 60~FPS~\cite{guo2024prediction, li2025unified, liu2024rdt, kim24openvla} so that the robot can respond smoothly.
However, existing AR-based VLA models introduce long inference latency on mobile platforms.
Here, we measure seven models~\cite{kim24openvla, wang2025unified, pertsch2025fast, goyal2025vla0, qu2025spatialvla, wang25vqvla, cen2025worldvla} on a Nvidia Thor SoC~\cite{thorsoc}.
In \Fig{fig:vla_latency}, most models can barely achieve 10~FPS, similar to numbers reported in prior work~\cite{kim2025fine, zhou2026characterizing}.
Thus, directly deploying VLA models on existing mobile platforms is insufficient for real-time execution.

\vspace{0pt}
\PBox{\textbf{Insight 2:} \textit{Decoding is time-consuming in VLA inference, leading to severe GPU underutilization.}}
\vspace{0pt}

\begin{figure}[t]
\centering
\begin{minipage}[t]{0.48\columnwidth}
  \centering
  \includegraphics[width=\columnwidth]{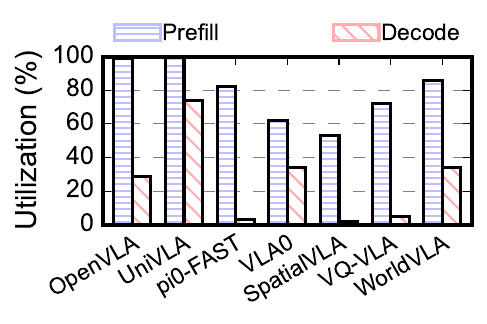}
  \caption{The tensor core utilization of the prefill and the decode stage. Tensor cores are severely underutilized during the decoding stage.}
  \label{fig:tc_util}
\end{minipage}
\hspace{2pt}
\begin{minipage}[t]{0.48\columnwidth}
  \centering
  \includegraphics[width=\columnwidth]{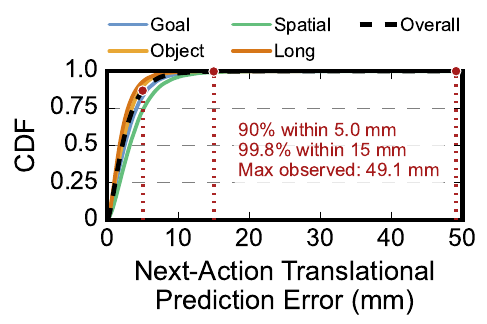}
  \caption{Action temporal correlation across four LIBERO suites~\cite{liu2023libero}. It shows strong short-term temporal correlation in robotic execution.}
  \label{fig:temporal_similarity}
\end{minipage}
\end{figure}

In \Sect{sec:bg:vla}, AR-based VLA models generate action tokens sequentially, limiting token-level parallelism.
Similar to LLMs, decoding is the key bottleneck in VLA inference, taking 69.5\% of the execution time on the LIBERO dataset~\cite{liu2023libero} (\Fig{fig:exec_breakdown}).
Meanwhile, on an Nvidia Thor SoC~\cite{thorsoc}, single-token decoding primarily involves matrix-vector operations, which underutilize tensor cores optimized for matrix-matrix multiplication.
Decoding exhibits lower tensor-core utilization than prefill (\Fig{fig:tc_util}).
Thus, increasing decoding parallelism is key to improving GPU utilization and inference performance.

\vspace{0pt}
\PBox{\textbf{Insight 3:} \textit{Robotic execution has strong temporal correlation that can be exploited to reduce inference computation.}}
\vspace{0pt}

Meanwhile, robot actions are not independent across time.
In high-frequency control, consecutive actions often change smoothly due to the continuous robot dynamics.
To quantify this temporal correlation, we analyze the traces from four LIBERO suites~\cite{liu2023libero}, covering over 2000 trajectories in \Fig{fig:temporal_similarity}.
For each sample, we linearly extrapolate the next action based on the previous two actions, and calculate the error between the extrapolated action and the actual action. 
We report the cumulative distribution function (CDF) of the translation error of our predictions. 
Overall, roughly 90\% of the next-action predictions have an error within 5.0~mm, which is within the operation error threshold~\cite{figureai, tenhumberg2020calibration}. 
This result shows a strong correlation in recent action history. 
Potentially, we can exploit this temporal correlation to reduce VLA computation.

\begin{figure*}
    \centering
    \includegraphics[width=\textwidth]{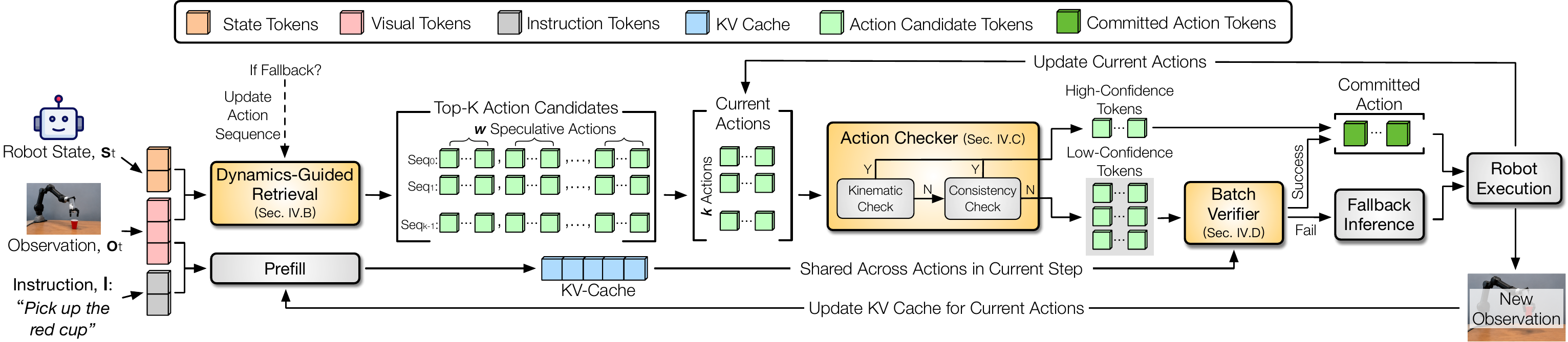}
    \caption{\proj overview.
    Given the current robot state $\mathbf{s}_t$, visual observation $\mathbf{o}_t$, and language instruction $\mathbf{I}$, \proj first encodes them into tokens in the latent space.
    The dynamics-guided retrieval module retrieves top-$k$ candidate action sequences with a length of $w$ from past execution contexts.
    An action checker then sequentially checks the retrieved action tokens and classifies them into high-confidence and low-confidence tokens.
    High-confidence tokens are directly committed for robotic execution, while low-confidence tokens are verified by the original VLA model in batches.
    Verified tokens are also committed for robot execution, whereas failed tokens trigger fallback inference to re-execute the original VLA model.
    After each action is executed, current actions are updated, and the KV cache is recomputed.
    The top-$k$ candidate sequences are refreshed whenever fallback is triggered, or candidate sequences are all verified.}
    
    \label{fig:framework_overview}
\end{figure*}

\section{\proj Framework}
\label{sec:algo}

To reduce the compute overhead of decoding in VLA models, we propose our framework, \proj, that exploits speculative decoding~\cite{chen2023accelerating, leviathan2023fast, miao2024specinfer} to verify multiple speculative action tokens in parallel and accelerate the end-to-end inference time.
We first present the overview of \proj in \Sect{sec:algo:overview}.
We then describe each key component of the framework in the following subsections.

\subsection{Overview}
\label{sec:algo:overview}

\paragraph{Intuition.}
The goal of our framework, \proj, is to reduce the compute overhead of decoding in AR-based VLA inference.
To do that, we exploit \textit{speculative decoding}.
Instead of invoking the original VLA model to decode every action token sequentially (\Fig{fig:ar_vla}), \proj proposes a lightweight prediction mechanism to draft a short sequence of future actions by exploiting robotic dynamics.
These drafted action tokens are then verified by the original VLA model in parallel to reduce the overall latency.
Next, we describe the workflow of \proj.

\paragraph{Workflow.}
\Fig{fig:framework_overview} gives an overview of our proposed framework, \proj.
At each control step, \proj takes three inputs: the visual observation $\mathbf{o}_t$, the robot state $\mathbf{s}_t$, and the language instruction $\mathbf{I}$.
These inputs are first processed by their corresponding encoders to obtain input embedding tokens.

Initially, before executing the VLA model, \proj invokes a lightweight drafting mechanism, called dynamics-guided (DG) retrieval (\Sect{sec:algo:search}), to predict future action tokens.
DG retrieval takes the robot state and visual observation as the query context and queries a retrieval database.
It then retrieves the top-$k$ action sequences with $w$ actions whose execution contexts are most similar to the current context.
The retrieved sequences are treated as action candidates and sent to an action checker.
Meanwhile, the model performs prefill in parallel to generate a KV cache, which is later used by a batch verifier.

Next, an action checker in \Sect{sec:algo:checker} leverages the robotic dynamics to determine which candidate actions need VLA model verification action-by-action.
Specifically, the action checker classifies candidate tokens within one action into high-confidence and low-confidence tokens.
High-confidence tokens are directly committed and executed by robots without VLA verification.
Low-confidence tokens are sent to the batch verifier, which checks whether they match the actions predicted by the original VLA model.
To further reduce verification overhead, we co-design an efficient batch verification in \Sect{sec:algo:verifier}.

If a low-confidence candidate token passes verification, it is directly used to control the robot.
Each time an action is executed, the current action is updated, and the KV cache is recomputed.
If the candidate token fails verification, the system then falls back to normal VLA inference.
When fallback inference is triggered, the entire action candidate sequences are also re-generated to refresh the current execution state.

\subsection{Dynamics-Guided Retrieval}
\label{sec:algo:search}

\paragraph{Intuition.}
The goal of DG retrieval is to generate accurate draft actions with low overhead for batched verification.
\Fig{fig:temporal_similarity} shows that short-term robot actions often exhibit strong temporal regularity.
Therefore, given the current visual observation and robot state, the near-future motions can be predicted from past execution contexts.
This naturally motivates us to leverage speculative decoding to accelerate VLA inference.

\paragraph{Why Retrieval-Based Drafting?}
Conventional speculative decoding techniques often rely on a small LLM draft model to generate speculative tokens.
However, this design has two limitations for robotic platforms.
First, this small draft model typically needs to be aligned with the original large VLA model to achieve a high acceptance rate.
Such alignment often requires model distillation~\cite{zhou2024distillspec, liu2024online} with additional training, which is time-consuming.
Second, even a small draft model can introduce non-trivial computation overhead for resource-constrained platforms.
To avoid these costs, \proj proposes a training-free retrieval-based drafting technique.

\begin{figure}
    \centering
    \includegraphics[width=\columnwidth]{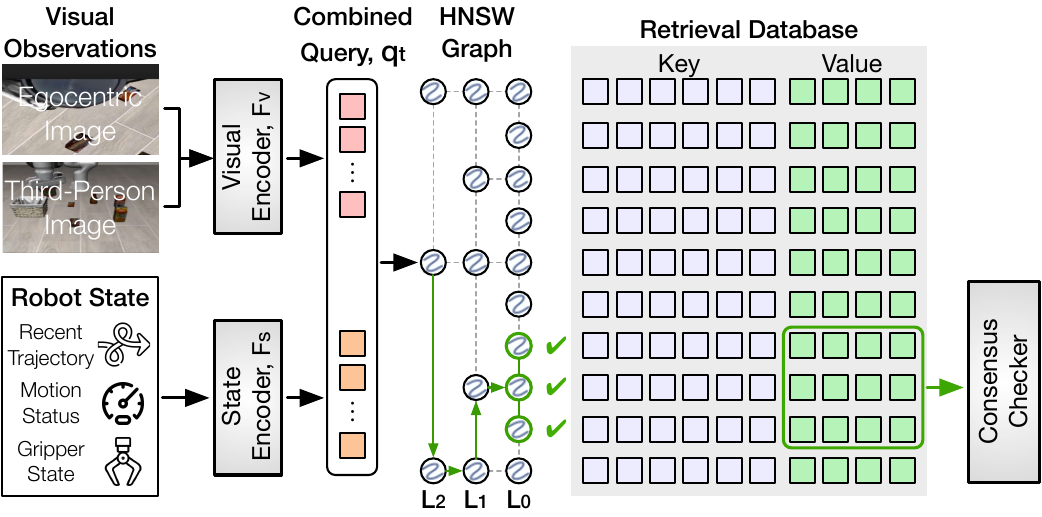}
    \caption{The process of DG retrieval.
    DG retrieval first constructs a query embedding from the current visual observations and robot state.
    DG retrieval then performs HNSW-based search over the retrieval database to find the top-$k$ keys that are most similar to the query.
    The associated values are returned as candidate action sequences and passed to the consensus checker.}
    \label{fig:dg_retrieval}
\end{figure}

\paragraph{Process.}
\Fig{fig:dg_retrieval} shows the overall process of our DG retrieval. 
Here, we use the visual observation $\mathbf{o}_t$ and the current robot state $\mathbf{s}_t$ to construct the retrieval query.
The visual observation includes an egocentric image captured by the robot-mounted camera and a third-person image captured by an external camera.
The robot state captures a short history of the robot execution state, including its recent pose trajectory, current motion status, and gripper state.
Together, these two inputs capture sufficient information for the current execution context.
Note that, we do not include the language instruction $\mathbf{I}$ in the input query.
Because different instructions sometimes lead to similar short-term trajectories.
Excluding the instruction allows DG retrieval to focus on robot dynamics and improves the generality of our technique.

Given $\mathbf{o}_t$ and $\mathbf{s}_t$, DG retrieval first generates a query embedding, which concatenates the embeddings of $\mathbf{o}_t$ and $\mathbf{s}_t$,
\begin{equation}
\mathbf{q}_t = \text{Concat}(\mathcal{F}_{v}(\mathbf{o}_t), \mathcal{F}_{s}(\mathbf{s}_t)), \nonumber
\end{equation}
where $\mathcal{F}_{v}$ and $\mathcal{F}_{s}$ denote the query encoder for visual observation and robot state, respectively.
The visual encoder is DINOv2~\cite{oquab2023dinov2}, the state encoder is normalization.
Given $\mathbf{q}_t$, we search through the retrieval database with similar keys.
In a retrieval database, each entry, $i$, stores a key-value pair.
The key is an embedding of a previous execution context, and the value is the future action trajectory that followed that context,
\begin{equation}
\mathbf{k}_i = \text{Concat}(\mathcal{F}_{v}(\mathbf{o}_i), \mathcal{F}_{s}(\mathbf{s}_i)),  \mathbf{v}_i = \{a_{i+1}, a_{i+2}, \ldots, a_{i+w}\}, \ \nonumber
\end{equation}
where $w$ is the length of the predicted action window. 

To efficiently search the retrieval database, DG retrieval uses an HNSW-based approximate nearest-neighbor search.
Given the query embedding $\mathbf{q}_t$, the HNSW search traverses through an HNSW graph as shown in \Fig{fig:dg_retrieval}.
In this example, after three layers' traversal, the HNSW search returns $\{\mathbf{v}_k\}$, a set of the top-$k$ database entries whose keys are closest to $\mathbf{q}_t$,
\begin{equation}
\{\mathbf{v}_k\} =
\operatorname{TopK}
\left(
\operatorname{sim}(\mathbf{q}_t, \mathbf{k}_i)
\right).
\end{equation}
In \Fig{fig:dg_retrieval}, we select the top-3 nodes highlighted in green.
In actual implementation, the default value of $k$ is 6.
`$\operatorname{sim}$' denotes the similarity metric used by the retrieval index.
Here, we use cosine similarity.
For each selected entry, DG retrieval returns the associated future action sequence.
The output of DG retrieval is a set of top-$k$ candidate trajectories.

We design DG retrieval to return a set of top-$k$ short trajectories for two reasons.
First, retrieving multiple trajectory candidates does not necessarily increase verification overhead; instead, it creates a verification batch that can better utilize tensor cores by verifying multiple candidates in parallel.
Second, multiple trajectory candidates provide additional information to estimate draft confidence.
Thus, action checker can directly commit reliable tokens and reduce the compute overhead. 

\subsection{Action Checker}
\label{sec:algo:checker}


Given $k$ sequences of drafted action candidates, the action checker identifies high-confidence actions that can be directly sent to robot execution.
Specifically, the action checker performs two complementary checks: a \textit{kinematic check} and a \textit{consistency check}.
The kinematic check evaluates whether a predicted action follows a smooth trajectory implied by the robot's recent motion history.
The consistency check evaluates whether the drafted action is consistently predicted across the top-$k$ drafted action candidates.
Actions that pass either check are regarded as high-confidence predictions and are accepted without invoking the subsequent model verification.
We next describe these two checks in the following paragraphs.

\paragraph{Kinematic Check.}
\Sect{sec:motiv} shows that 90\% of the actions can be extrapolated linearly.
The kinematic check leverages this property to avoid compute-intensive model verification when a drafted action follows the robot’s recent motion trend.

Specifically, we propose a continuity score, $\cS_c$, to measure how well an action $\mathbf{a}_i$ follows the extrapolated trajectory,
\begin{equation}
\cS_c(\mathbf{a}_i) = \min\left(\frac{1}{\alpha}\cF_{\Delta}(\mathbf{a}_i, \mathbf{a}'_i), 1\right),
\end{equation}
where $\mathbf{a}'_i$ denotes the action extrapolated for timestep $i$ from the previous two actions, $\cF_{\Delta}$ measures the discrepancy between $\mathbf{a}_i$ and $\mathbf{a}'_i$, and $\alpha$ is a normalization constant with a default value of 0.5.
The function $\cF_{\Delta}$ is defined as,
\begin{align}
    \cF_{\Delta}(\mathbf{a}_{i}, \mathbf{a}'_{i}) & = \|\Delta x_i - \Delta x'_i\|_2 + \|\Delta r_i- \Delta r'_i\|_2 + \phi (g_i, g'_i), \nonumber \\
    \text{and}  &\ \phi (g_i, g'_i) = \begin{cases}
       1,\ \text{if}\ g_i\neq g'_i, \\
       0,\ \text{otherwise},        \\
    \end{cases}
\end{align}
\par\phantomsection\label{rev:rcq3}\par\phantomsection\label{rev:rcq4}
where $\Delta x_i$, $\Delta r_i$, and $g_i$ denote the normalized displacement, rotation difference, and gripper state of action $\mathbf{a}_i$, respectively.
Thus, $\cF_{\Delta}$ captures the combined discrepancy across the translational, rotational, and gripper components of the predicted and extrapolated actions.
The extrapolated action $\mathbf{a}'_i$ is computed from the actions at previous two timesteps as,
\begin{equation}
\mathbf{a}'_i =
\{(2\Delta x_{i-1} - \Delta x_{i-2}),
(2\Delta r_{i-1} - \Delta r_{i-2}),
g_{i-1}\}.
\end{equation}

To evaluate the continuity of a candidate sequence, $\mathbf{L}_{\mathbf{a}} = \{\mathbf{a}_{t+1}, \ldots, \mathbf{a}_{t+w}\}$, obtained from DG retrieval, we compare the continuity score of each action against a kinematic threshold $\theta_\text{kin}$.
If $\cS_c(\mathbf{a}_i)$ is lower than $\theta_\text{kin}$, the action is considered kinematically consistent and directly becomes the committed action.
Otherwise, this action in the current step undergoes the consistency check and, if that fails, batch verification.

Here, $\theta_\text{kin}$ is not a fixed value.
Instead, $\theta_\text{kin}$ is dynamically adjusted according to the smoothness of the recent action history $\mathbf{L}_\text{his} = \{\mathbf{a}_{t-h+1}, \ldots, \mathbf{a}_{t}\}$.
$h$ is the length of the recent history, with a default value of 16.
We first measure the continuity of recent actions as,
\begin{equation}
\cS_\text{his} = \frac{1}{h}\sum_{\mathbf{a} \in \mathbf{L}_\text{his}} \cS_c(\mathbf{a}).
\end{equation}
The kinematic threshold, $\theta_\text{kin}$, is then computed as
\begin{equation}
\theta_\text{kin} = \min\left(\max\left(
{\beta - \cS_\text{his}}, \theta_\text{kin,min} \right), \theta_\text{kin,max} \right),
\end{equation}
where $\beta$ is a hyperparameter, 0.5. 
$\theta_\text{kin,min}$ and $\theta_\text{kin,max}$ denote the lower and upper bounds of $\theta_\text{kin}$, respectively.
They are 0.20 and 0.45, respectively.
When $\cS_\text{his}$ is small, the recent action history is smoother, so we increase $\theta_\text{kin}$ to allow more drafted actions to pass the kinematic check.
Conversely, when $\cS_\text{his}$ is large, the recent motion is less smooth; thus, we decrease $\theta_\text{kin}$ and become more conservative.
When multiple candidate action tokens in the current step satisfy the threshold $\theta_\text{kin}$, we choose the action with the lowest overall continuity score.

\paragraph{Consistency Check.}
The kinematic check allows some drafted actions to skip batch verification by exploiting the smoothness of robot dynamics.
However, some actions may not satisfy the kinematic smoothness criterion, yet can still be reliable if they are consistently predicted across multiple candidate sequences retrieved by DG retrieval.
To cover those actions, we propose a consistency check that further identifies drafted actions that can be accepted without batch verification.

Given $k$ candidate action sequences, we first compute the weighted mean action $\bar{\mathbf{a}}_i$ at each timestep $i$:
\begin{equation}
\bar{\mathbf{a}}_i =
\frac{\sum_{j=1}^{k} w^j \mathbf{a}_{i}^{j}}
{\sum_{j=1}^{k} w^j}, \quad w^j = \exp({s^j})
\end{equation}
\par\phantomsection\label{rev:rdq2}
where $w^j$ denotes the confidence of the $j$-th candidate sequence obtained from DG retrieval, $s^j$ is the cosine similarity score between the top-$k$ candidate embedding and the query embedding.
$\mathbf{a}_{i}^{j}$ denotes the $i$-th action in that sequence.
We then select the action closest to $\bar{\mathbf{a}}_i$ among the top-$k$ candidates as the consensus action:
\begin{equation}
\mathbf{a}_{i,\text{mid}} =
\underset{{\mathbf{a}_{i}^{j}, j \in {1,\ldots,k}}}{\operatorname{argmin}}
\cF{\Delta}(\mathbf{a}_{i}^{j}, \bar{\mathbf{a}}_i).
\end{equation}

Next, we measure how consistently the top-$k$ candidate sequences support the action $\mathbf{a}_{i,\text{mid}}$ using a consistency score:
\begin{equation}
\cS_{i,\text{con}} =
\frac{\sum_{j=1}^{k} w^j \cF_{\Delta}(\mathbf{a}_{i,\text{mid}}, \mathbf{a}_{i}^{j})}
{\sum_{j=1}^{k} w^j}.
\end{equation}
A lower $\cS_{i, \text{con}}$ means stronger consensus among the candidate actions at timestep $i$.
If $\cS_{i, \text{con}}$ is below a predefined consistency threshold $ \theta_\text{con} = 0.2$, this consensus action is high confidence, and it can be directly committed to the committed action queue.
Otherwise, the action is considered uncertain, and the $k$ actions in the current step are sent for batch verification.

\subsection{Batch Verifier}
\label{sec:algo:verifier}

While the action checker allows some drafted actions to skip model verification, the remaining uncommitted actions still need to be verified by the original VLA model.
A naive approach is to verify these actions one by one, but this introduces redundant data movement, e.g., repeatedly loading KV cache.
To reduce this overhead, we propose a mask-based attention mechanism that shares the KV cache across different action candidates during verification.

\begin{figure}[t]
\centering
\begin{minipage}[t]{0.48\columnwidth}
  \centering
    \includegraphics[width=\columnwidth]{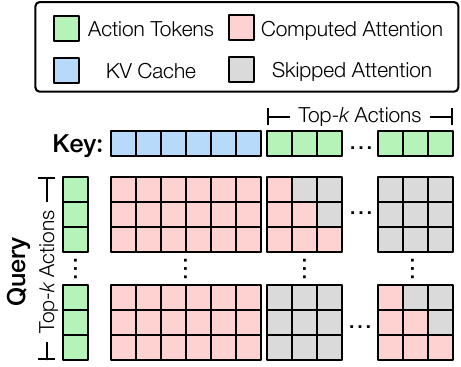}
    \caption{Masked attention for batched verification. Multiple top-$k$ action candidates are packed into a single attention computation and share the same prefix KV cache.
    }
    \label{fig:masked_attn}
\end{minipage}
\hspace{2pt}
\begin{minipage}[t]{0.48\columnwidth}
  \centering
    \includegraphics[width=\columnwidth]{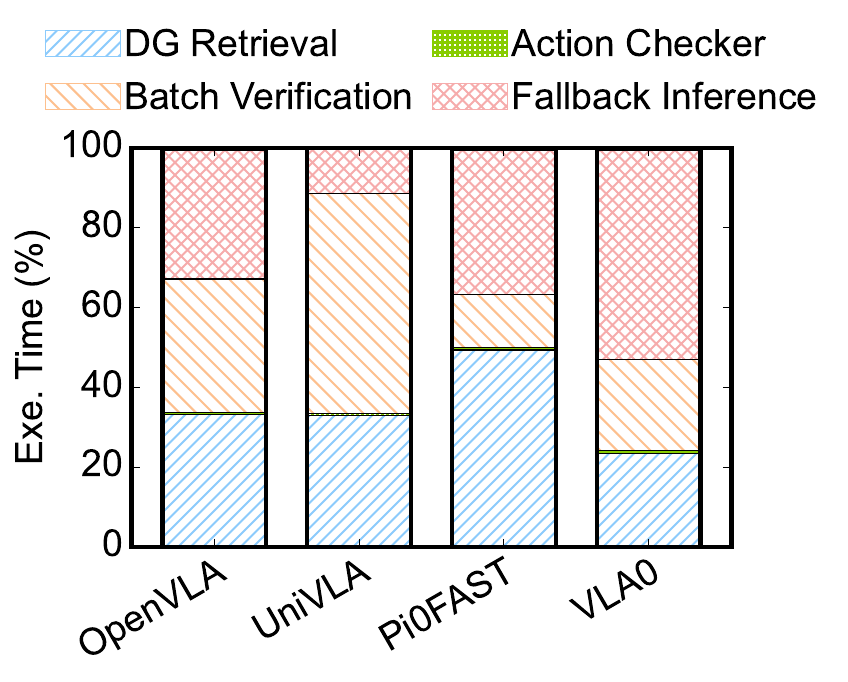}
    \caption{The execution breakdown after applying our software optimizations. DG retrieval becomes one of the bottlenecks, taking 35\% of execution time across four models.}
    \label{fig:issue}
\end{minipage}
\end{figure}

\paragraph{Attention.}
\Fig{fig:masked_attn} shows our masked attention mechanism used for batched verification.
Instead of verifying each candidate sequence independently, we pack the top-$k$ actions in the current step into a single attention computation.
These candidates share the same prefix context; thus, we only load their KV cache once and share it across all candidates.

To preserve the independence of different candidate branches, we apply an attention mask: each action token can attend to the shared KV cache and the preceding tokens within the same candidate sequence, but cannot attend to tokens from other candidates, as shown in \Fig{fig:masked_attn}.
As a result, the verifier computes only the necessary branch-specific attention while skipping invalid cross-candidate attention.
This way, we avoid repeated KV-cache loading during batch verification.

Our mechanism mainly targets the attention computation and KV-cache reuse.
It does not optimize the linear operators, which have been extensively studied in prior acceleration work~\cite{kim2025ditto, kong2024cambricon, heo2025exion}.
Therefore, our approach is orthogonal to existing optimizations for linear projections.

\paragraph{Verification.}
Specifically, for each verified timestep, we compare each verified candidate action with its corresponding drafted action and compute their discrepancy score using $\cF_{\Delta}$.
The candidate with the lowest discrepancy score is selected.
We then compare this score against a predefined acceptance threshold $\theta_\text{ver}$, which is 20.
If the score is below $\theta_\text{ver}$, the selected action is accepted and committed to the action queue.
Otherwise, the system falls back to the normal VLA inference utilizing the current observation, and the prefilled prefix KV cache generated by the verifier may be reused.
\par\phantomsection\label{rev:cq1}

\subsection{Novelty of \proj}
\label{sec:algo:novelty}

All prior works~\cite{wang2025spec, zheng2026kerv, zheng2026heisd} still follow the same speculative decoding execution, i.e., model drafting followed by batch verification. 
Their key contributions lie in techniques proposed to improve draft acceptance rates.
For instance, HeiSD~\cite{zheng2026heisd} leverages both RAG retrieval and draft-based models to generate draft tokens.
KERV~\cite{zheng2026kerv} uses robotic kinematics for acceptance threshold adjustment.
SpecVLA~\cite{wang2025spec} directly adopts the off-the-shelf speculative decoding.
These methods have to rely on target-model predictions, and they are often model-specific designs with explicit finetuning.

\proj instead proposes a \textit{unified and portable} approach that uses execution experience for both draft prediction and confidence estimation.
\textcircled{\scriptsize 1}
\proj expands reusable action knowledge across multiple VLA models and proposes a RAG-based draft predictor to retrieve multiple short-term trajectories.
\textcircled{\scriptsize 2}
A lightweight checker jointly exploits recent-motion continuity and retrieved-trajectory agreement to directly commit high-confidence actions and selectively verify uncertain candidates.
\textcircled{\scriptsize 3}
Masked attention shares the KV context across candidate sequences and verifies them in parallel, while preserving independence between candidates.


\section{Architectural Support}
\label{sec:arch}

We first motivate the need for architectural support in \Sect{sec:arch:issue}.
We then present an overview of the proposed architecture in \Sect{sec:arch:overview}.
Lastly, we explain two key components: the \proj core in \Sect{sec:arch:acc} and the checker unit in \Sect{sec:arch:checker}.

\subsection{Issues}
\label{sec:arch:issue}

\Fig{fig:issue} shows the execution breakdown after applying these optimizations.
Although the average decoding latency is reduced by 64.9\% (see \Sect{sec:eval:gpu}), DG retrieval becomes one of the bottlenecks and contributes non-trivial latency to the decoding stage.
Moreover, HNSW search involves irregular graph traversal, leading to irregular memory accesses that are inefficient on mobile GPUs.
Therefore, we propose a principled yet minimal set of hardware augmentations on existing DNN accelerators to support HNSW-based DG retrieval.

\begin{figure}
    \centering
    \includegraphics[width=\columnwidth]{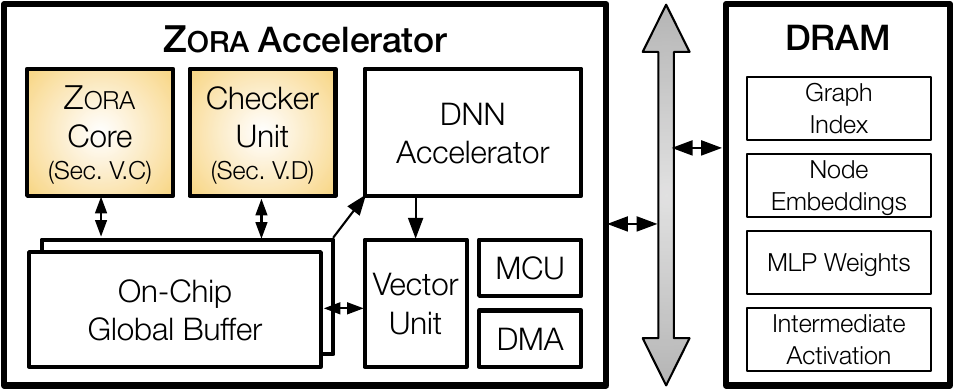}
    \caption{Overview of the \proj accelerator architecture. The accelerator contains a \proj core to accelerate DG retrieval, a checker unit, and a DNN accelerator. Yellow blocks highlight our key contributions.}
    \label{fig:arch}
\end{figure}

\begin{figure*}[t]
    \centering
    \includegraphics[width=\textwidth]{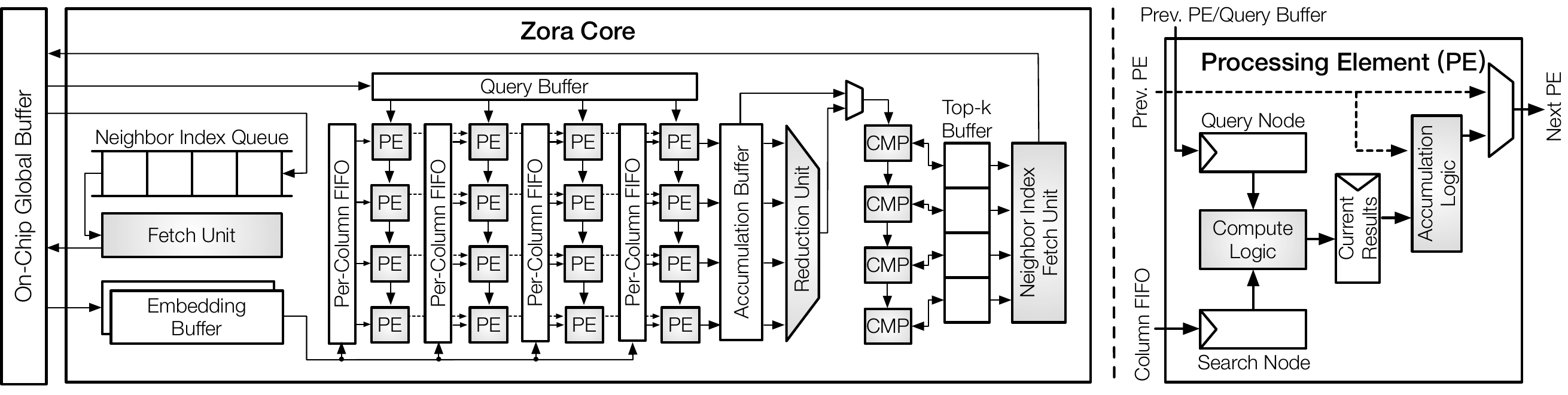}
    \caption{Overview of the \proj core architecture. The core accelerates HNSW-based DG retrieval through a four-stage pipeline: fetch, compute, accumulation, and update. The PE array exploits \textit{node-level parallelism} and \textit{embedding-level parallelism} to compute partial similarity scores between the query embedding and neighbor embeddings. Partial similarity scores are accumulated and then consumed by either the reduction unit for greedy search in upper HNSW layers or the comparison units for top-$k$ selection in the bottom layer. Lastly, the neighbor index fetch unit updates the search frontier for unvisited nodes.}
    \label{fig:core}
\end{figure*}

\subsection{Overview}
\label{sec:arch:overview}

To address the issue described in \Sect{sec:arch:issue}, we propose our \proj architecture.
We show its overview in \Fig{fig:arch}.
The \proj accelerator augments a DNN accelerator with two major components: a \proj core and a checker unit.

During execution, the \proj core first accelerates DG retrieval to predict future pose trajectories, while the checker unit performs the operations required by the action checker.
A microcontroller then orchestrates the overall execution flow and performs lightweight control operations.
Lastly, a DNN accelerator accelerates batch verification of the retrieved pose trajectories by exploiting their value similarity.
The DNN accelerator can be implemented as a systolic array or another design; \proj is not tied to a specific DNN accelerator architecture.
\Sect{sec:eval:hw} shows that our design can be integrated with different DNN accelerators.
Meanwhile, a vector unit is also included to support vector operations.

DRAM stores the input and output data for the \proj core and the DNN accelerator, e.g., the graph index for HNSW traversal, node embeddings, model weights, and intermediate activations.
The accelerator also includes a double-buffered on-chip buffer to overlap computation with data fetching.

\subsection{\proj Core}
\label{sec:arch:acc}

\paragraph{Design Principle.}
The goal of the \proj core is to accelerate HNSW-based DG retrieval.
Although HNSW search is inherently sequential due to the graph traversal nature, our architecture exploits two additional sources of parallelism: \textit{node-level parallelism} and \textit{embedding-level parallelism}.
First, for each node, HNSW needs to evaluate multiple neighboring nodes and identify the neighbor closest to the query.
These neighbor evaluations are independent and can be processed in parallel.
We call it node-level parallelism (NLP).
Second, computing the similarity between the query embedding and a node embedding requires computing across all embedding dimensions.
These dimensions can be partitioned into multiple groups and processed in parallel.
We call it embedding-level parallelism (ELP).
The overall design of the \proj core is built upon these two forms of parallelism.

\paragraph{Overview.}
\Fig{fig:core} gives an overview of the \proj core, which still performs HNSW search layer-by-layer.
The overall execution consists of four pipelined stages: \textit{fetch}, \textit{compute}, \textit{accumulation}, and \textit{update}.
The fetch stage loads the embeddings of candidate neighbor nodes.
The compute stage then calculates the partial similarity between the query embedding and each candidate node embedding.
The accumulation stage combines partial similarity scores into full similarity scores.
Lastly, the update stage selects the nearest node or top-$k$ nodes and updates the search frontier for the next traversal loop.
We further explain these four stages in the following paragraphs.

\paragraph{Fetch.}
The fetch stage is responsible for loading candidate neighbor nodes for similarity evaluation.
As shown in \Fig{fig:core}, the neighbor index queue stores the indices of unvisited neighboring nodes.
At each traversal step, the fetch unit dequeues a batch of node indices from the neighbor index queue and fetches their embeddings from the on-chip global buffer or off-chip memory.
The fetched embeddings are stored in the embedding buffer before being consumed by the compute stage.
To overlap irregular embedding fetching with similarity computation, the embedding buffer is double-buffered.

\paragraph{Compute.}
The compute stage calculates the similarity between the query embedding and candidate node embeddings.
The query embedding is first loaded into the query buffer and then broadcast to the PE array.
Candidate node embeddings are loaded from the embedding buffer into per-column FIFOs.

The PE array is designed to exploit both NLP and ELP.
Each row is assigned to one candidate neighbor node, while each column is assigned to one embedding channel group, i.e., a subset of embedding dimensions.
Therefore, different rows process different candidate nodes in parallel, and different PEs within the same row compute partial similarity scores of a single embedding over different dimensions.

Each column of PEs operates like a one-dimensional systolic array.
The query embedding values are reused within one column, while different PEs in a column read the values of different candidate nodes.
The right side of \Fig{fig:core} shows the PE design.
Each PE receives a query embedding element from the previous PE and a candidate-node embedding element from the per-column FIFO.
It then performs the similarity computation and stores partial accumulated similarity scores in local registers.
Depending on the similarity metric, the compute logic can support different operations.

\paragraph{Accumulation.}
After each PE finishes computing the partial similarity score for its assigned embedding-channel group, the partial score is forwarded to the next PE in the same row via the inter-PE forwarding lane, shown by the dashed path in \Fig{fig:core}.
During forwarding, the accumulation logic combines partial scores across columns and obtains the full similarity scores, which are stored in the accumulation buffer.

There are two ways to consume these scores, depending on the HNSW layer.
For the upper layers, HNSW performs greedy search and only needs to identify the single closest neighbor.
Therefore, the full similarity scores are sent to a reduction unit, which selects the nearest node among the evaluated candidates.
For the bottom layer, on the other hand, HNSW returns the top-$k$ nearest neighbors.
In this case, the full similarity scores are directly read from the accumulation buffer and streamed into an array of comparison units.
Each score is compared against the current top-$k$ candidates and forwarded along the comparison array.
If a candidate node is closer than an existing entry in the top-$k$ buffer, the comparison unit updates the top-$k$ buffer and sends an update signal to the neighbor-index fetch unit.

\paragraph{Update.}
The update stage determines how the HNSW traversal proceeds.
If the current layer is an upper layer, the nearest node selected by the reduction unit becomes the entry point for the next search step.
If the nearest node is closer to the query than the current node, the neighbor index fetch unit is updated to fetch the unvisited neighbors of this new node.
If no neighbor is closer, the search proceeds to the next layer.

At the bottom layer, the update stage maintains the top-$k$ buffer.
When a newly evaluated node is closer than one of the existing top-$k$ candidates, the top-$k$ buffer is updated.
The neighbors of newly selected candidates are then inserted into the neighbor index queue if they have not been visited.
This is done via the on-chip lookup buffer.
This process continues until there is no new index in the neighbor index queue.
Thus, we find the top-$k$ nearest embedding nodes.

\begin{figure}
    \centering
    \includegraphics[width=\columnwidth]{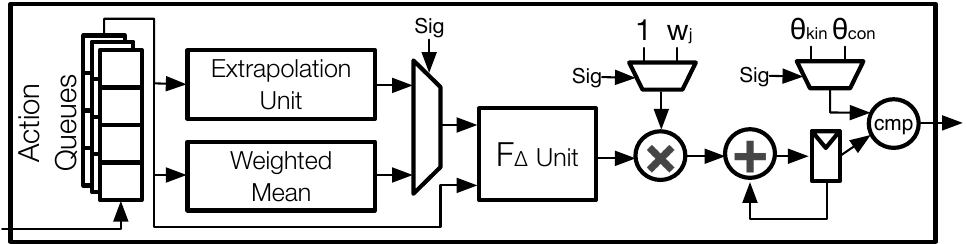}
    \caption{Design of the checker unit. It computes kinematic or consistency discrepancy scores with $\cF_{\Delta}$ and compares them with thresholds to decide whether to commit an action or send it for model verification.}
    \label{fig:checker_unit}
\end{figure}

\subsection{Checker Unit}
\label{sec:arch:checker}

\Fig{fig:checker_unit} shows the design of the checker unit, which supports both the kinematic check and the consistency check described in \Sect{sec:algo:checker}.
The action queue buffers recent actions and drafted candidate actions.
For the kinematic check, the extrapolation unit predicts the next action from the recent action history.
For the consistency check, the weighted-mean unit computes the consensus action across the top-$k$ candidates.
A control signal selects which action is sent to the $\cF_{\Delta}$ unit, which computes the discrepancy between the drafted action and the selected action.
The resulting score then can be weighted by the candidate confidence $w^j$ and accumulated across candidates or action components.
Finally, the checker compares the accumulated score against the corresponding threshold to determine whether the action can be directly committed or should be sent for model verification.
\section{Experimental Setup}
\label{sec:exp}

\paragraph{Hardware Setup.}
The DNN accelerator in the base architecture is implemented as a systolic array with $192 \times 192$ PEs clocked at 1~GHz.
Each PE is designed for 16-bit fixed-point computation.
The architecture also includes a scalar unit with 192 parallel lanes.
The double-buffered global buffer is implemented using SRAM with 8~MB.
The \proj core contains a $4 \times 4$ PE array with an embedding buffer of 128~KB.
One checker unit is also included in our design.
We primarily evaluate \proj using this configuration.

\par\phantomsection\label{rev:rcq6}
\paragraph{Experimental Methodology.}
The RTL implementations of the main \proj components are synthesized using Synopsys tools and placed and routed using Cadence tools in TSMC 16~nm FinFET technology.
The SRAM is generated using an ARM memory compiler.
Power is estimated with Synopsys PrimeTimePX using fully annotated switching activity.
Off-chip memory is modeled as eight Micron 32~Gb LPDDR5X at 8533~MT/s~\cite{lpddr5x} to match the DRAM bandwidth of Nvidia Thor SoC~\cite{thorsoc}, and DRAM energy is estimated using Micron's System Power Calculator~\cite{microdrampower}.
To evaluate system-level performance, we build a cycle-level simulator of the \proj architecture, with the latency and energy of each component parameterized using post-synthesis results.
We directly time the GPU execution as well as the kernel launch via CUDA events on Nvidia's Thor SoC under 1.53~GHz, 90W power mode, and BF16 precision.
The GPU power is directly measured using the built-in power sensing circuitry.

\paragraph{Area.}
The total area of \proj is 14.5~mm$^2$ at 16~nm.
Compared to a base systolic array with 8~MB on-chip buffer, our augmentation introduces a minimal overhead of 0.4\%.
Within \proj core, the $4\times4$ PE array and embedding buffer take 6.7\% and 88.3\%, respectively.
The rest takes 5.0\%.

\par\phantomsection\label{rev:raq6}

\paragraph{Software Setup.}
We evaluate \proj using four representative AR-based VLA algorithms with different action representations and decoding schemes: OpenVLA~\cite{kim24openvla}, UniVLA~\cite{wang2025unified}, Pi0FAST~\cite{pertsch2025fast}, and VLA0~\cite{goyal2025vla0}.
To evaluate our framework, we use a widely used LIBERO dataset~\cite{liu2023libero}.
Success rate is measured as the accuracy metric using the standard procedure.
Each VLA, we average the results over 4 suites, each including 10 tasks with 50 episodes, i.e., 2000 episodes per model.
\par\phantomsection\label{rev:cq5}
\par\phantomsection\label{rev:rcq5}
\par\phantomsection\label{rev:rcq8}


\paragraph{Retrieval Database.}
To construct the global database, we execute each original VLA model to collect successful trajectories.
For each VLA model, we perturb the official LIBERO initial states with random offsets to 50\% of the official state range.
For each task, we use separate seeds and collect around 50 successful traces with their metadata.
Database collection and evaluation use disjoint seeds and initial states; no evaluation episode is included in the database.
Overall, it contains a total of 7479 trajectories, with over 1.10$\times10^6$ data points.
We configure HNSW with at most 16 edges per node, retrieve ($k$=6) candidates, and set the action-sequence length to ($w$=6).

\paragraph{Variants.}
To show that \proj can be integrated with different accelerators, we compare the following baselines:
\begin{itemize}
\item \mode{GPU}: a mobile Blackwell GPU in Nvidia Thor SoC~\cite{thorsoc}.
\item \mode{\proj+GPU}: our software-version acceleration framework running on the GPU in the Nvidia Thor SoC~\cite{thorsoc}.
\item \mode{Base+SA}: a baseline accelerator with a DNN accelerator implemented as a systolic array. It runs the baseline algorithm without our speculative decoding acceleration.
\item \mode{\proj+SA}: our framework with speculative decoding and a systolic array as the DNN accelerator.
\item \mode{Base+Ditto}: a baseline accelerator with a DNN accelerator implemented using Ditto~\cite{kim2025ditto}. 
It leverages the temporal similarity across frames to accelerate the prefill stage and runs the baseline algorithm without our speculative decoding acceleration.
\item \mode{\proj+Ditto}: our accelerator with a Ditto-like~\cite{kim2025ditto} accelerator and our speculative decoding acceleration.
\end{itemize}
\par\phantomsection\label{rev:rdq5}
For both \mode{Base+Ditto} and \mode{\proj+Ditto}, Ditto is applied only to the prefill stage.
Applying it to decoding requires caching and accessing linear-layer outputs from the previous iteration, introducing additional memory traffic that exacerbates the memory-bound decoding bottleneck.

\section{Evaluation}
\label{sec:eval}

\subsection{Accuracy}
\label{sec:eval:acc}

\begin{figure}[t]
    \centering
    \includegraphics[width=\columnwidth]{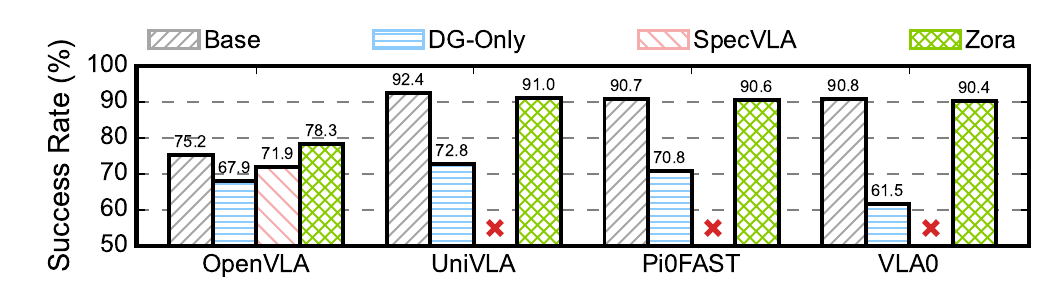}
    \caption{Accuracy comparison over different speculative methods. Red crosses show \mbox{\mode{SpecVLA}} is unavailable for the corresponding models.}
    \label{fig:eval_accuracy}
\end{figure}

\Fig{fig:eval_accuracy} shows the accuracy comparison among different methods.
Here, we compare three methods: 1) \mode{DG-Only}, which only uses the results from DG retrieval without any verification; 2) \mode{SpecVLA}~\cite{wang2025spec}, which uses a small DNN as a draft model; 3) \mode{\proj}, which is our full implementation.
\mode{Base} is the original algorithm without speculative decoding.

\par\phantomsection\label{rev:raq7}
\par\phantomsection\label{rev:raq8}

Overall, \mode{\proj} achieves similar accuracy to the original algorithm.
In contrast, \mode{SpecVLA} and \mode{DG-Only} drop the accuracy by 3.3\% and 19.0\%, respectively.
Meanwhile, the draft model in \mode{SpecVLA} needs to be retrained for each model; thus, we only show its result on OpenVLA.
\mode{\proj} sometimes achieves better accuracy than \mode{Base} potentially for two reasons.
First, our RAG database is comprehensive and incorporates traces from all VLA models.
Second, its kinematic and consistency checks further exploit robot dynamics, rather than solely relying on model inference.
Overall, \proj provides the best accuracy among existing speculative approaches with great portability.

\begin{figure}[t]
\centering
\begin{minipage}{0.68\columnwidth}
    \centering
    \includegraphics[width=\linewidth]{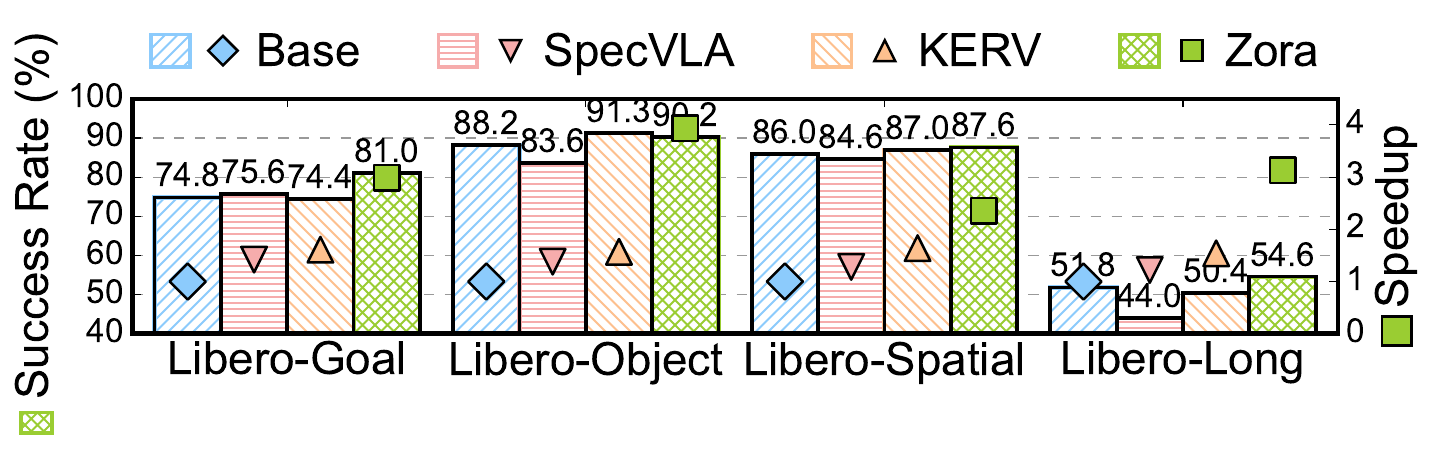}
    \caption{Accuracy and GPU performance comparison with \mode{SpecVLA~\cite{wang2025spec}} and \mode{KERV}~\cite{zheng2026kerv} on OpenVLA.
    Numbers are measured on a Thor SoC and normalized to \mode{Base}.}
    \label{fig:eval_comparsion}
\end{minipage}
\hfill
\begin{minipage}{0.30\columnwidth}
    \centering
    \includegraphics[width=\linewidth]{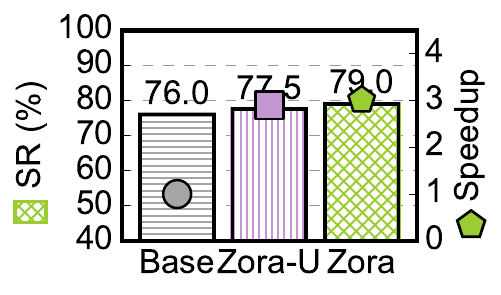}
    \caption{Unseen task evaluation over OpenVLA. \mode{\proj-U} shows the result on unseen held-out tasks.}
    \label{fig:eval_unseen}
\end{minipage}
\end{figure}

In \Fig{fig:eval_comparsion}, we also compare against two prior works, \mode{SpecVLA}~\cite{wang2025spec} and \mode{KERV}~\cite{zheng2026kerv}, over OpenVLA on the LIBERO dataset.
Both methods only target OpenVLA, not the other models.
Overall, \mode{\proj} achieves 8.9\% and 3.3\% higher accuracy compared to \mode{SpecVLA} and \mode{KERV}, respectively.
Meanwhile, \mode{\proj} has a much higher GPU speedup.

\par\phantomsection\label{rev:cq4}

\paragraph{Unseen Tasks.}
To evaluate the generalization of \proj to unseen tasks, we introduce \mode{\proj-U} in \Fig{fig:eval_unseen}.
Specifically, we partition the LIBERO tasks into separate 80\%/20\% subsets based on task category.
We then use the 80\% subset to construct its RAG database following \Sect{sec:exp}.
All methods in \Fig{fig:eval_unseen} are evaluated exclusively on identical episodes from the held-out 20\% tasks.
However, \mode{\proj} uses the full-task database.
\mode{\proj-U} achieves a success rate comparable to that of \proj, with only 1.5\% decrease.

\begin{figure}[t]
    \centering
    \includegraphics[width=\columnwidth]{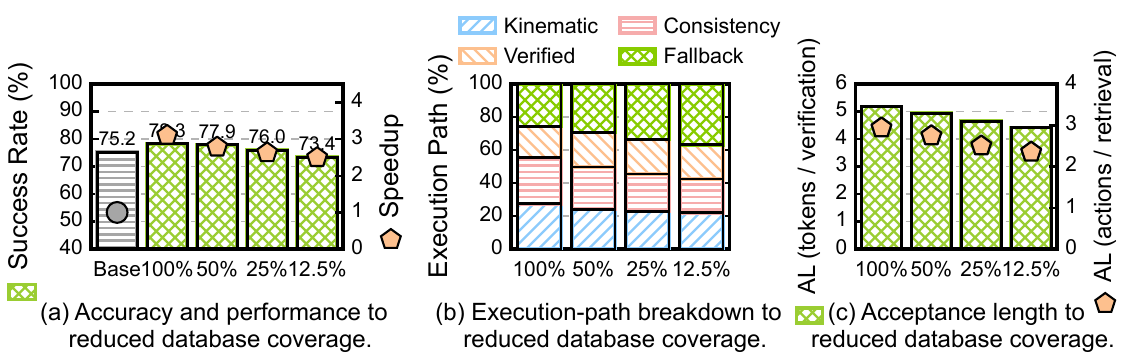}
    \caption{Robustness evaluation on \proj over different dataset coverage. All results are measured on \mbox{OpenVLA}. We show (a) accuracy and GPU performance, (b) execution path breakdown, and (c) acceptance length (AL).}  
    \label{fig:eval_robustness}
\end{figure}

\par\phantomsection\label{rev:cq3}
\par\phantomsection\label{rev:cq6}

\paragraph{Robustness.}
\Fig{fig:eval_robustness} evaluates \proj on OpenVLA as database coverage decreases from 100\% to 12.5\% by removing trace data.
\Fig{fig:eval_robustness}a shows that accuracy decreases from 78.3\% to 73.4\%, while GPU speedup also decreases gradually.
\Fig{fig:eval_robustness}b further shows the breakdown of different action commit paths.
With full database coverage, our kinematic and consistency checks can skip almost 60\% of model verification.
Lower coverage reduces direct acceptance through kinematic and consistency checks.
More actions rely on model verification and fallback inference. 
Lastly, in \Fig{fig:eval_robustness}c, both the number of accepted tokens per verification and actions per retrieval decline moderately, as database coverage decreases.

\subsection{GPU Performance}
\label{sec:eval:gpu}

\begin{figure}[t]
    \centering
    \includegraphics[width=\columnwidth]{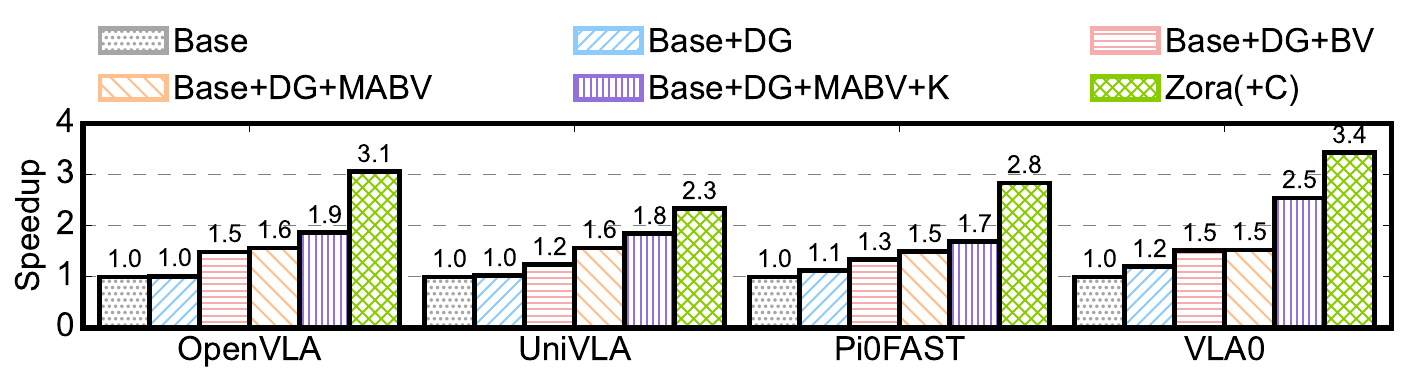}
    \caption{GPU performance of \proj variants. \mode{Base+DG} applies only DG retrieval; \mode{Base+DG+BV} applies conventional batch verification; \mode{Base+DG+MABV} further applies masked attention for batched verification with shared prefix KV cache; \mode{Base+DG+MABV+K} includes Kinematic Check; the full-fledged design, \mode{Zora(+C)}, includes Consistency Check.}
    \label{fig:gpu_speedup}
\end{figure}


\mbox{\Fig{fig:gpu_speedup}} shows the GPU performance of \mbox{\proj} with different variants of our design.
Here, \mbox{\mode{Base+DG}} applies only DG retrieval;
\mbox{\mode{Base+DG+BV}} further applies conventional batch verification. 
\mbox{\mode{Base+DG+MABV}} instead applies batch verification in \mbox{\Sect{sec:algo:verifier}}; \mbox{\mode{Base+DG+MABV+K}} further adds kinematic check on \mbox{\mode{Base+DG+MABV}};
\mbox{\mode{Zora(+C)}} is the full-fledged design with the \textit{Consistency Check} in \mbox{\Sect{sec:algo:checker}}.

\par\phantomsection\label{rev:rcq2}

In \Fig{fig:gpu_speedup}, on average, \mode{Base+DG} only achieves 1.1$\times$, because it still sequentially verifies each action candidate sequence.
\mode{Base+DG+BV} improves the speedup to 1.4$\times$ via conventional batch verification.
By combining with our batch verification, \mode{Base+DG+MABV} improves the performance to 1.6$\times$.
\mode{Base+DG+MABV+K} reaches 2.0$\times$ speedup by directly accepting dynamically reliable actions without model verification.
By further exploiting the robotic kinematics and consistency, \mode{Zora(+C)} eventually can achieve 2.9$\times$ speedup.
Thus, incorporating robot kinematic priors into speculative decoding can largely reduce the unnecessary verification.

Meanwhile, \Fig{fig:eval_comparsion} compares the \proj GPU performance against prior work~\cite{wang2025spec, zheng2026kerv} on a single Thor SoC.
On average, \mode{SpecVLA} and \mode{KERV} can 1.3$\times$ and 1.6$\times$ speedup, while \mode{\proj} can achieve a much higher speedup, 3.1$\times$.


\subsection{Hardware Performance}
\label{sec:eval:hw}


\begin{figure}[t]
    \centering
    \includegraphics[width=\columnwidth]{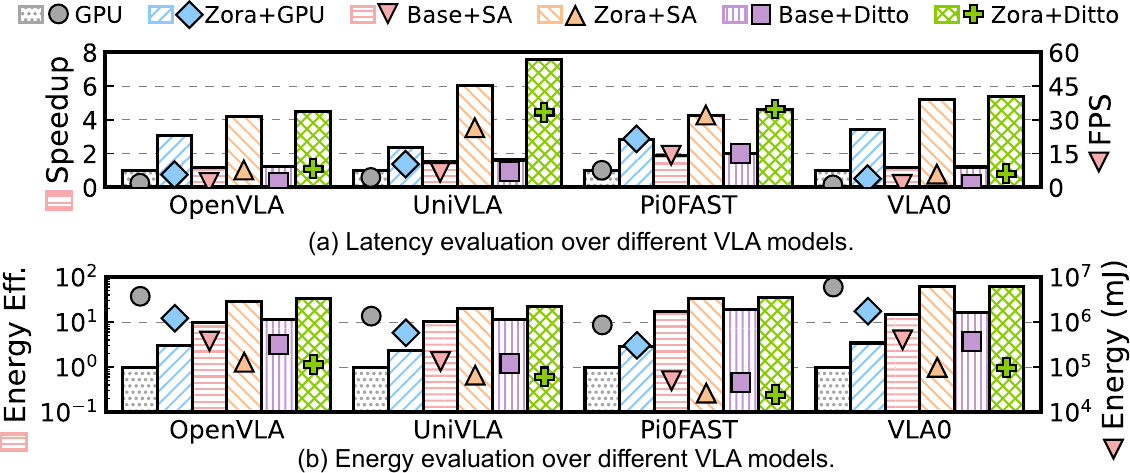}
    \caption{The performance and energy savings over different hardware variants. Bars show relative gains over the per-model \mode{GPU} baseline, and markers show absolute measurements, i.e., FPS and energy per task. 
    } 
    \label{fig:hw_perf}
\end{figure}

\par\phantomsection\label{rev:rcq7}
\paragraph{Speedup.}
\Fig{fig:hw_perf}a shows the performance comparison of \proj against different accelerators.
Here, all numbers are normalized to the corresponding \mode{GPU} baselines.
Here, we show two different accelerator baselines, \mode{Base+SA} and \mode{Base+Ditto}.
\mode{Base+SA} implements the DNN accelerator via a systolic array, while \mode{Base+Ditto} uses a dedicated accelerator, Ditto~\cite{kim2025ditto}.
Overall, \mode{Base+SA} and \mode{Base+Ditto} achieve only 1.4$\times$ and 1.5$\times$ speedup, respectively.
This is because the decoding stage is primarily memory-bound.


On the other hand, by integrating our framework with our hardware augmentation, 
\mode{\proj+SA} and \mode{\proj+Ditto} achieve 4.9$\times$ and 5.5$\times$ speedup, respectively.
There are three main reasons leading to higher speedup.
First, the \proj framework reduces the overall verification workload.
Second, the batch verification makes the decoding stage less memory-bound.
Third, our hardware augmentation drastically reduces the DG retrieval overhead.
Overall, the results show that \proj has great compatibility with DNN accelerator designs and consistently achieves high speedup.

In terms of frame rate, \mode{\proj+Ditto} achieves 34.8~FPS and 33.4~FPS on Pi0FAST and UniVLA, respectively.
That said, decoding in \proj remains memory-bound, which leads to the \proj FPS being around 30~FPS, e.g., on UniVLA.
On the other hand, prefill and DG retrieval are compute-bound, with a peak bandwidth requirement of 182.9~GB/s.
Prefill and DG retrieval take 150.9 and 32.0~GB/s, respectively. 
The scalability analysis in \Sect{sec:eval:scal} further shows \proj can achieve real-time (60FPS) with higher bandwidth.

\begin{figure}[t]
    \centering
    \begin{minipage}[t]{0.48\columnwidth}
        \includegraphics[width=\columnwidth]{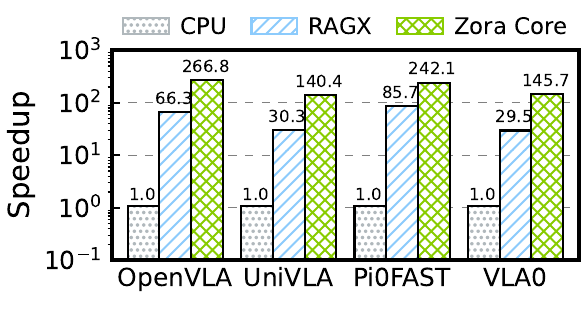}
        \caption{DG retrieval speedup of the \mbox{\proj} core over the CPU baseline on Thor SoC and RAGX~\cite{mahapatra2025in}.}
        \label{fig:retrieval_speedup}
    \end{minipage}
    \hfill
    \begin{minipage}[t]{0.48\columnwidth}
        \includegraphics[width=\columnwidth]{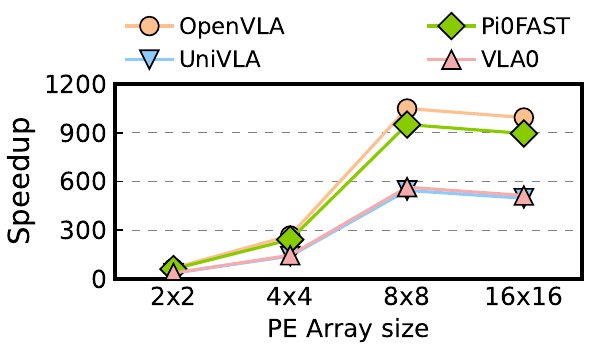}
        \caption{The scalability of the DG retrieval speedup to PE array size.}
        \label{fig:retrieval_scala}
    \end{minipage}
\end{figure}

Since DG retrieval is one of the bottlenecks before acceleration, we further isolate the retrieval performance in \Fig{fig:retrieval_speedup}.
\Fig{fig:retrieval_speedup} compares the performance of the \proj core over the CPU baseline and RAGX~\cite{mahapatra2025in}, a RAG accelerator for LLM serving.
The CPU baseline is measured on an Arm Neoverse V3AE CPU on a Thor SoC.
RAGX~\cite{mahapatra2025in} is also configured in a way with the same number of PEs and buffer sizes.
On average, \proj core achieves 198.8$\times$ speedup over the corresponding CPU baseline and 3.7$\times$ speedup over RAGX under the same hardware resource constraints.
The speedup over RAGX is largely from the fine-grained node-level and embedding-level parallelism proposed in \Sect{sec:arch:acc}.

\paragraph{Energy Savings.}
\Fig{fig:hw_perf}b shows the energy savings of \proj against different hardware baselines.
Here, we define energy savings as the ratio between the energy consumption of \mode{GPU} and that of the corresponding accelerator.
The overall trend aligns with the speedup results.
For instance, \mode{Base+SA} achieves 13.1$\times$ energy efficiency over the GPU baseline, and \mode{\proj+SA} improves this to 35.8$\times$.
Similarly, \mode{\proj+Ditto} also achieves higher energy efficiency over \mode{Base+Ditto}.
Even without hardware support, \mode{\proj+GPU} also delivers 2.9$\times$ energy efficiency over \mode{GPU}.

We also report energy per task in \Fig{fig:hw_perf}b.
We accumulate the energy across all the inference within one task.
On average, the energy is reduced from 2.9$\times10^6$~mJ (\mode{GPU}) to 7.4$\times10^4$~mJ (\mode{\proj+Ditto}).
The average power is reduced from 40.0~W (\mode{GPU}) to 6.5~W (\mode{\proj+SA}) and 6.9~W (\mode{\proj+Ditto}).


\subsection{Scalability Study}
\label{sec:eval:scal}

\begin{figure}
    \centering
    \begin{minipage}[t]{0.48\columnwidth}
        \includegraphics[width=\columnwidth]{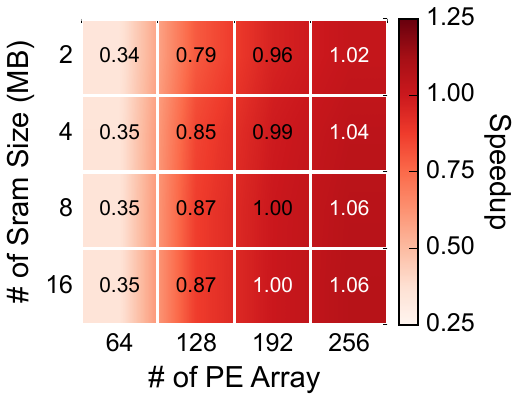}
        \caption{Scalability of the DNN accelerator speedup to PE array size and SRAM capacity.}
        \label{fig:dnn_scalability}
    \end{minipage}
    \hfill
    \begin{minipage}[t]{0.48\columnwidth}
        \includegraphics[width=\columnwidth]{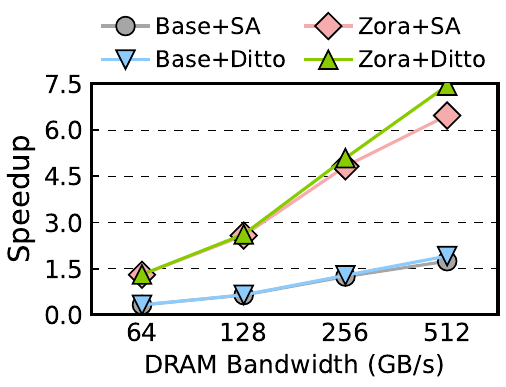}
        \caption{Impact of DRAM bandwidth on end-to-end speedup.}
        \label{fig:dram_scalability}
    \end{minipage}
\end{figure}

\begin{figure*}[t]
    \centering
    \includegraphics[width=\textwidth]{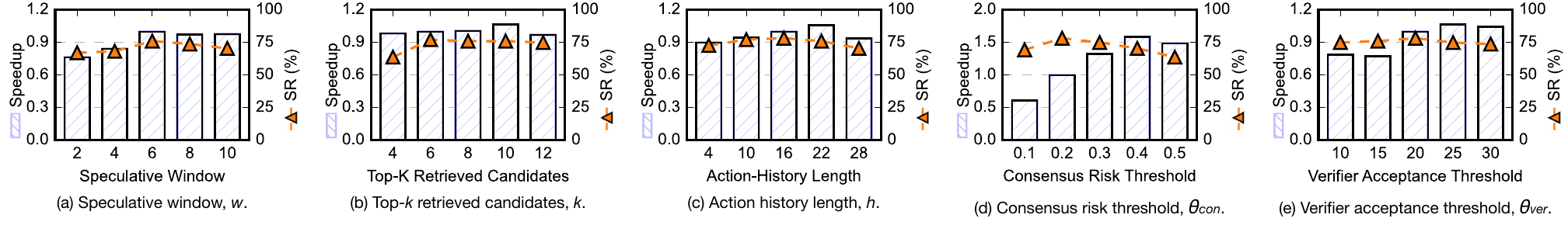}
    \caption{Sensitivity of performance and task success rate to different algorithmic hyperparameters in the \proj framework.}
    \label{fig:sens_study}
\end{figure*}

\paragraph{\proj Core.}
\Fig{fig:retrieval_scala} shows the impact of the PE size in \proj core on retrieval speedup.
The default configuration, 4$\times$4, achieves 198.8$\times$ speedup over the CPU baseline.
This reduces the retrieval latency to $<$1\% of the execution time.
Increasing the PE array to 8$\times$8 further improves the retrieval speedup to 776.9$\times$.
However, it provides negligible end-to-end speedup benefit.
Further scaling to 16$\times$16 yields no additional gain, as the workload shifts to memory-bound.

\paragraph{DNN Accelerator.}
\Fig{fig:dnn_scalability} shows how the PE array size of the DNN accelerator and on-chip buffer size affect the overall performance.
Here, we show the scalability of the systolic array via a case study of OpenVLA.
Overall, the on-chip buffer has a small impact on the overall performance.
As for PE array size, performance scales near-linearly from 64$\times$64 to 192$\times$192.
Further scaling to 256$\times$256 PEs yields less performance improvement.

\paragraph{DRAM Bandwidth.}
\Fig{fig:dram_scalability} shows the impact of DRAM bandwidth on the speedup across the four hardware variants.
Increasing DRAM bandwidth improves the performance of all variants, as VLA decoding is mainly memory-bound.
For example, \mode{\proj+SA} improves from 1.3$\times$ speedup at 64~GB/s to 6.5$\times$ speedup at 512~GB/s.
Compared with baseline designs, \proj benefits more from increased bandwidth because its batch verification alleviates the memory requirements.

\subsection{Sensitivity Study}
\label{sec:eval:sens}

\Fig{fig:sens_study} shows the sensitivity of hardware performance and task success rate to different hyperparameters in the \proj framework mentioned in \Sect{sec:algo}.
All experiments are conducted on OpenVLA as case studies.
The performance numbers are normalized to the default configurations.

\paragraph{Speculative Window, $w$.}
\Fig{fig:sens_study}a shows the performance and accuracy sensitivity to the speculative window, $w$.
The default value is 6.
$w = 6$ achieves the highest accuracy and speedup.
Further increasing the speculative window will not improve the performance because fewer actions are accepted.

\paragraph{Top-$k$ Retrieved Candidates, $k$.}
\Fig{fig:sens_study}b shows the performance and accuracy sensitivity to Top-$k$ retrieved candidates, $k$.
The default value is 6.
Further increasing $k$ can potentially improve the performance.
However, a larger $k$ might also introduce less accurate candidates and affect task accuracy.
Selecting $k$ between 6 and 10 provides a good balance.

\paragraph{Action History Length, $h$.}
\Fig{fig:sens_study}c shows the performance and accuracy sensitivity to the action history length, $h$.
The default value is 16.
Increasing $h$ provides more historical information and improves the effectiveness of the kinematic check.
However, an excessively long history introduces outdated motion patterns and reduces the accuracy of action prediction.
We find that $h=16$ has a good trade-off.

\paragraph{Consensus Risk Threshold, $\theta_\text{con}$.}
\Fig{fig:sens_study}d shows the sensitivity to the consensus risk threshold $\theta_\text{con}$.
The default value is 0.2.
A smaller threshold often makes the consensus check more conservative, leading to lower speedup.
Meanwhile, a larger threshold allows more actions to skip verification; setting $\theta_\text{con}$ too high may introduce inaccurate actions.
The accuracy degradation at 0.3 is due to the threshold-based verification-skipping mechanism.
Increasing this threshold leads to more candidates bypassing verification.
Actions near the acceptance boundary may switch to unverified execution.

\paragraph{Verifier Acceptance Threshold, $\theta_\text{ver}$.}
\Fig{fig:sens_study}e shows how the verifier acceptance threshold $\theta_\text{ver}$ affects the performance and accuracy.
The default value is 20.
A lower threshold requires stronger agreement between drafted and verified actions.
This reduces the number of accepted actions and performance gains.
We find that $\theta_\text{ver}$ between 20 and 25 provides a good trade-off between performance and accuracy.

\section{Related Work}
\label{sec:related}

\paragraph{Acceleration for Robotics.}
A broad range of accelerators and systems have been proposed to improve the efficiency of robotics workloads over the past decade~\cite{sacks2018robox, hadidi2021quantifying, liu2021micro, neuman2021robomorphic, krishnan2023autopilot, bakhshalipour2022racod, yang2023dadu, shah2023energy, neuman2023roboshape, nikiforov2023rose, shah2024collison, hao2024orianna, bakhshalipour2024tartan, choe2025ana, huang2025corki, ting2025hiper, zhang2025idea, chen2025octocache, wan2025reca}.
Early works primarily targeted specific robotic kernels or operations~\cite{sacks2018robox, bakhshalipour2022racod, yang2023dadu, liu2021micro, shah2023energy, shah2024collison}, by exploiting their domain-specific computation patterns.
Later efforts shifted toward system-level acceleration, optimizing common bottlenecks across multiple robotics algorithms rather than individual kernels~\cite{nikiforov2023rose, hao2024orianna, krishnan2023autopilot}.
Some studies also leverage 3D Gaussian splatting for SLAM localization~\cite{huang2026splatonic, li2025rtgs}.
More recently, architectural research has expanded toward embodied AI~\cite{huang2025corki, wan2025reca} and autonomous driving~\cite{chen2025octocache, choe2025ana}, where performance increasingly depends on end-to-end system stacks.
In contrast, \proj targets the unique challenges of AR-based VLA inference by co-designing speculative decoding algorithms with dedicated hardware support.

\par\phantomsection\label{rev:raq3}
\paragraph{Speculative Decoding.}
Speculative decoding~\cite{leviathan2023fast, chen2023accelerating, miao2024specinfer, cai2024medusa} has been widely studied to accelerate the decoding stage in LLM inference. 
Early approaches~\cite{leviathan2023fast, chen2023accelerating} introduce a lightweight draft model to generate candidate tokens.
Other system-level optimizations~\cite{miao2024specinfer, cai2024medusa} further optimize the batch verification efficiency. 
REST~\cite{he2024rest} accelerates LLM through retrieval-based drafting and verifying tokens with the target model.
While \proj adopts a similar workflow, the key of \proj is to establish a dynamics-guided retrieval system and a selective action verification procedure.

Building upon these advances, recent works~\cite{wang2025spec, zheng2026kerv, zheng2026heisd, huo2025spec} have explored speculative decoding to reduce the inference latency for AR-based VLA models. 
For instance, Spec-LLaVA~\cite{huo2025spec} investigates different strategies for speculating future tokens.
Others~\cite{wang2025spec, zheng2026kerv} also further optimize the draft-and-verify process in the speculation process to improve action prediction accuracy.
Our work addresses this challenge by introducing RAG-augmented and kinematics-aware speculation and provides more accurate action generation while improving the overall performance.
Meanwhile, \proj introduces a hardware augmentation to further boost the performance with negligible area overhead.
\par\phantomsection\label{rev:cq2}

\paragraph{RAG Acceleration.}
Prior work accelerates retrieval via near-memory processing~\cite{quinn2025compute, quinn2025accelerating}, in-memory/storage processing~\cite{mahapatra2025in, zhang2025characterizing}.
These studies target server-scale RAG workloads, whereas \proj targets the latency of individual action-retrieval queries in VLA inference.
In particular, IKS~\cite{quinn2025accelerating} accelerates exact nearest-neighbor search, while RAGX~\cite{mahapatra2025in} supports diverse retrieval algorithms and embedding generation.
\proj specializes its retrieval engine by exploiting node-level parallelism across candidate neighbors and embedding-level parallelism within similarity calculations to unlock new parallelism without requiring query batching.
\section{Conclusion}
\label{sec:conc}

Overall, \proj demonstrates that efficient VLA inference requires jointly exploiting
the characteristics of VLA workloads and the intrinsic dynamics of robotic systems.
\proj achieves up to 3.4$\times$ speedup with software-only optimization and up to 7.6$\times$ speedup with minimal hardware augmentation.
Our results show that VLA model workloads can be effectively optimized by incorporating domain-specific robotic knowledge.
This co-design provides a promising direction for building efficient, real-time embodied AI systems.


\bibliographystyle{IEEEtranS}
\bibliography{refs}

\end{document}